# Turbulence, Turbulent Mixing, and Gravitational Structure Formation in the Early Universe

Carl H. Gibson[1]

[1]Mechanical and Aerospace Engineering and Scripps Institution of Oceanography Departments, University of California at San Diego,
La Jolla, CA, 92093-0411, USA, cgibson@ucsd.edu, http://www-acs.ucsd.edu/~ir118

**ABSTRACT**
Turbulence and turbulent mixing of temperature powered the big bang formation of the universe at Planck length, time, and temperature scales. Planck-Kerr inertial-vortex forces balanced Planck gravitational forces to produce Planck (particle-pair) gas, Planck-gas turbulence and space-time-energy. Inflation at the strong force temperature fossilized the turbulence. Gluon-neutrino-photon bulk viscous forces exceeded gravitational and inertial-vortex forces of the baryonic (ordinary) matter during the radiation dominated hot plasma epoch, and large diffusive velocities of the weakly interacting nonbaryonic dark matter (possibly neutrinos) prevented gravitational structure formation of this material, contrary to the 1902 Jeans acoustic criterion and "cold dark matter" models. The first plasma structures were proto-galaxy-super-cluster fragments and voids triggered by turbulent-viscous-gravitational masses matching the horizon mass at time $10^{12}$ s. The last plasma structures formed were proto-galaxies. At the $10^{13}$ s plasma-gas transition, the proto-galaxies fragmented to form proto-globular-star-cluster clumps of earth-mass primordial-fog-particles (PFPs) that now comprise the baryonic dark matter. Observational evidence of PFPs is provided by young globular star clusters formed when galaxies merge, thousands of cometary knots seen around dying stars, and the rapid twinkling of lensed quasar images.

**1. INTRODUCTION**
When was the first turbulence in the universe? When was the first turbulent mixing? When was the first gravitational structure formation? For the first time it may be possible to answer such questions. New information about the early universe is flooding in from a variety of very high resolution telescopes on the earth surface and in space, covering frequency bands previously unavailable. A new generation of computers and software are struggling to keep pace, along with a new generation of theoreticians and theories. Cosmological and astrophysical models for gravitational structure formation that remained undisturbed for decades during the last century are now confronted daily with new data that must be accommodated or the models abandoned.

The standard model of gravitational structure formation is presented in books about cosmology such as Weinberg 1972, Silk 1989, Kolb and Turner 1990, Peebles 1993, Padmanabhan 1993 and Rees 2000. Unfortunately, the model is based on the gravitational instability criterion of Jeans 1902, where density fluctuations on scales larger than the Jeans length scale $L_J = V_S/(\rho G)^{1/2}$ are unstable to gravitational condensation and those smaller are not, where $V_S$ is the speed of sound and G is Newton's constant of gravity. The Jeans theory and Jeans gravitational condensation criterion are disputed by Gibson 1996, 2000 and elsewhere. Alternative length scale criteria for gravitational instability are proposed depending on viscous, turbulent and diffusive constraints. These significantly alter the predicted times and mass scales of gravitational structure formation. The standard Jeans-CDM model is discussed in Section 2, the Gibson 1996-2002 model in Section 3, comparisons to observations in Section 4, and conclusions in Section 5.

**2. STANDARD COSMOLOGICAL MODEL FOR STRUCTURE FORMATION**
Jeans 1902 applied linear perturbation theory to the inviscid Euler equations with gravity, reducing the fluid mechanical problem to one of gravitational acoustics. Sound waves travel a wave length $\lambda$ in a time period $\tau_S = \lambda/V_S$. The gravitational free fall time $\tau_{FF}$ is $(\rho G)^{-1/2}$. The two time scales $\tau_S$ and $\tau_{FF}$ are equal for $\lambda = L_J$. Short wavelength sound waves propagate away before they are affected by gravity. Long wavelength sound waves are gravitationally unstable. Jeans' criterion for gravitational structure formation is that density fluctuations on scales $L > L_J$ are unstable and $L < L_J$ are stable as long as L is less than the horizon scale $L_H$.

Jeans' theory ignores effects of viscous and turbulent forces on gravitational structure formation, as well as diffusivity. Most density maxima and minima do not propagate with the sound speed, but diffuse to a symmetric configuration and then move with the fluid. On scales smaller than $L_J$, such non-acoustic density maxima and minima are absolutely unstable to gravitational forces in the absence of viscous and turbulence forces or strong diffusivity D. The diffusion velocity D/L matches the gravitational velocity $L(\rho G)^{1/2}$ at the Schwarz diffusive length scale $L_{SD} = [D^2/\rho G]^{1/4}$. Self gravitational forces match viscous forces at the viscous Schwarz scale $L_{SV} = [\gamma\nu/\rho G]^{1/2}$, and gravity matches turbulence forces at the Schwarz turbulent scale $L_{ST} = \varepsilon^{1/2}/[\rho G]^{3/4}$, where $\rho$ is the gas density, G is $6.72 \times 10^{-11}$ $m^3$ $kg^{-1}$ $s^{-2}$, $\gamma$ is the rate-of-strain, and $\varepsilon$ is the viscous dissipation rate. Pressure forces are in acoustical equilibrium on scales smaller than $L_J$, and cannot prevent either gravitational collapse on density maxim, or prevent expansion as gravitational rarefaction waves to form voids at density minima. Pressure support and thermal support are misconceptions deriving from the invalid Jeans 1902 instability criterion. The maximum speed of rarefaction waves is the sound speed $V_S$, from well known gas dynamics, and this limits the time scale of void formation rather than the free fall time $\tau_{FF}$. The Gibson 1996, 2000 criterion for gravitational structure formation is that density fluctuations are unstable on scales $L > [L_{SD}, L_{ST}, L_{SV}]_{Max}$, as long as $L < L_H$. As soon as the largest Schwarz scale is less than the horizon scale $L_H$, structure will form, even if $L_J > L_H$.

Because the sound speed $V_S$ in the plasma epoch before $10^{13}$ s (300,000 years) is $c/3^{1/3}$, the Jeans scale $L_J$ is always larger than the horizon scale of causal connection $L_H = ct$, where c is the speed of light and t is the time since the big bang. Therefore no gravitational structure can form in the primordial baryonic plasma of electrons, protons and alpha particles ($^4$He).  Even after the hot gas forms, the Jeans mass $M_J = \rho L_J^3$ is about $10^{36}$ kg, or a million solar masses.  Without fragmentation at smaller scales or some nonbaryonic magic, such massive gas blobs would rapidly collapse to form short-lived superstars, supernovas and massive black holes in brief periods (a few million years); that is, no planets, no stars, no galaxies, no galaxy-clusters, and no galaxy-super-clusters could ever appear.  The Jeans model fails immediately for a Universe consisting entirely of baryons.

To salvage the observed Universe within Jeans' theory it is necessary to hypothesize the existence of "cold dark matter" (CDM).  The sound speed $V_S$ is $[kT/m]^{1/2}$ for an ideal gas consisting of particles of mass m at temperature T, where $k = 1.38 \; 10^{-23}$ kg m$^2$ s$^{-2}$ K$^{-1}$ is Boltzmann's constant.  It has been known for many years that the universe is nearly flat and that a huge amount of invisible (dark) matter must exist in galaxies and galaxy clusters to prevent them from flying apart due to centrifugal forces.  Most of the invisible matter must be nonbaryonic because the baryonic matter density is limited to < 3% by precise constraints of neucleosynthesis, Weinberg 1972.  A flat universe today has critical density $\rho_{Crit} = 10^{-26}$ kg m$^{-3}$ so its expansion coasts gradually toward zero with time rather than accelerating for a $\rho < \rho_{Crit}$ open universe, or reversing toward a big crunch for a $\rho > \rho_{Crit}$ closed universe.  The observed $\rho$ is indistinguishable from and remarkably close to the flat-universe critical value (some claim differently), which means that about 99.9% of the matter is dark, about 97% of the dark matter is nonbaryonic, and about 97% of the baryonic matter is also dark.  The standard model assumes the nonbaryonic matter is also "cold" with $L_J < L_H$ during the plasma epoch so it can condense to provide gravitational potential wells to capture the baryonic matter and cause it to form the stars and galaxies observed.  Why is this imaginary "cold-dark-matter" cold?  The answer is that it must be cold or Jeans' theory won't work.  But how can it be cold if it is collisionless and was formed hot?  An expanding ideal gas cools because it does pressure work on the expanding surroundings at the expense of its internal energy, but a collisionless gas can't exert pressure and can't do any such work, so its particles must maintain their velocities and temperatures forever.  Figure 1 gives an outline of the standard CDM cosmological model and the large number of questionable areas for the evolution of gravitational structure by this scenario.

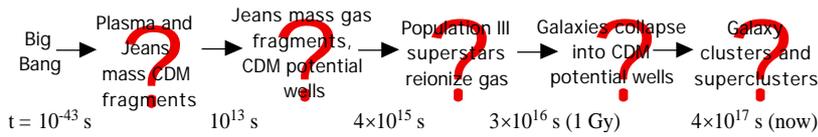

**Figure 1.  The standard model for structure formation is highly questionable.**

As shown in Fig. 1, somehow the nonbaryonic matter must form massive CDM seeds during the plasma epoch to nucleate the larger scale structures observed.  The seeds cluster and begin to collect gas by gravity, circumventing the Jeans criterion for the hot gas and triggering the formation of a (tailored) number of superstars with masses comparable to the seed masses; that is, $10^3$ to $10^6$ solar masses ($M_{SUN} = 2 \; 10^{30}$ kg).  If the cold dark matter (CDM) is so weakly collisional that it can't be observed, then yet another misconception is needed to prevent the CDM potential wells from diffusing away.  This is provided by Binney and Tremaine 1987, p190, where it shown that gravitationally bound systems of weakly collisional particles have relaxation times R/v N/lnN, where R/v is the crossing time for the CDM protogalaxies, R is the size, v is the particle velocity and N is the number of particles. To produce a flat universe with kinetic energy precisely matching gravitational potential energy using neutrinos (the only known form of non-baryonic dark matter) the neutrino mass m must be about $10^{-35}$ kg. Thus N/lnN is about $10^{75}$ for a neutrino CDM potential well of dwarf galaxy mass $10^{39}$ kg, giving relaxation times for reasonable values of R/v. The situation is no different for more massive neutralinos, another CDM candidate (there are dozens), where the relaxation time is about $10^{77}$ s. The problem with this model is that it neglects the phenomena of core collapse, Binney and Tremaine 1987, p528. For any initial condition of the CDM blob, all of its particles pass near the core in a free fall period $\tau_{FF}$ and the core becomes progressively more concentrated with time constant $\tau_{FF}$. No matter how small the collision cross section of the CDM particles, they are forced to interact at the core and randomise their velocities in times of order $\tau_{FF}$. The effective diffusivity of the blob $R^2/\tau_{FF}$ monotonically increases toward the particle diffusivity $D = L_C v$ and R approaches $L_{SD}$, Gibson 2002. CDM seed fragments cannot form in the first place, but if they were to they would rapidly diffuse away and not cluster.

Further difficulties beset the standard model. Quasars are very bright sources of light at great distances, with maximum observed red shifts greater than 6. Very little quasar light is absorbed compared to that expected from the hydrogen budget (the Gunn-Peterson paradox), so it is assumed in the standard model that the hydrogen has been re-ionized. The re-ionization of the universe requires superstars and supernovas, with feedback, capable of reversing the plasma to gas transition, Ciardi et al. 2000. Unfortunately for CDM, the extremely violent events and huge amount of energy required to re-ionize all the free hydrogen and helium gas would have had detectable side effects, such as the formation of strong turbulence, contrary to the laminar universe indicated by the small $\Delta T/T \approx 10^{-5}$ cosmic microwave background T fluctuations, and a large production by Population III supernovae of heavy elements not seen in the metal-free Population II stars of globular star clusters in the outskirts of galaxies. Tailored CDM clumping is needed to limit the number of such superstars. The dense concentrations of small, ancient stars in globular clusters observed could not possibly have formed under conditions of strong Population III supernova turbulence because the large $L_{ST}$ scales of the turbulence would have prevented it, just as it does in spiral galaxy disks that contain mostly widely spaced metal-rich population I stars like the sun, and open star clusters like the Pleiades. In the Gibson 1996 model, Population III stars never existed and the universe never reionized.

A major difficulty for the standard CDM model is the timing of large structure formation, which occurs very late. The imaginary CDM protogalaxies take a billion years or so to collect enough baryons to form collapsing galaxies of stars, the galaxies take a lot of time to form clusters, and the clusters take a lot of time to form superclusters. Unfortunately, very high resolution telescopes on the ground and in space can now see mature galaxies in clusters and superclusters at times when none of these structures should exist by the standard CDM model, Steidel et al. 2000. N body CDM numerical simulations show lots of galaxies should still be present in the $10^{24}$ m supervoids between the $10^{47}$ kg superclusters, but these spaces are found to be quite empty, Peebles 2001. Observations of supervoid densities show values about $10^{-35}$ kg m$^{-3}$, corresponding to no more than a dozen globular clusters and no galaxies at all in a supervoid. Peebles suggests voids pose a crisis for CDM models. The crisis may require abandoning Jeans-CDM entirely. Without Jeans-CDM, extremely empty supervoids are expected since these are the first gravitational structures to form starting in the first 10% of the plasma epoch at $10^{12}$ s. Figure 2 shows more details of the standard CDM model based on Fig. 1 of Ostriker and Gneden 1996.

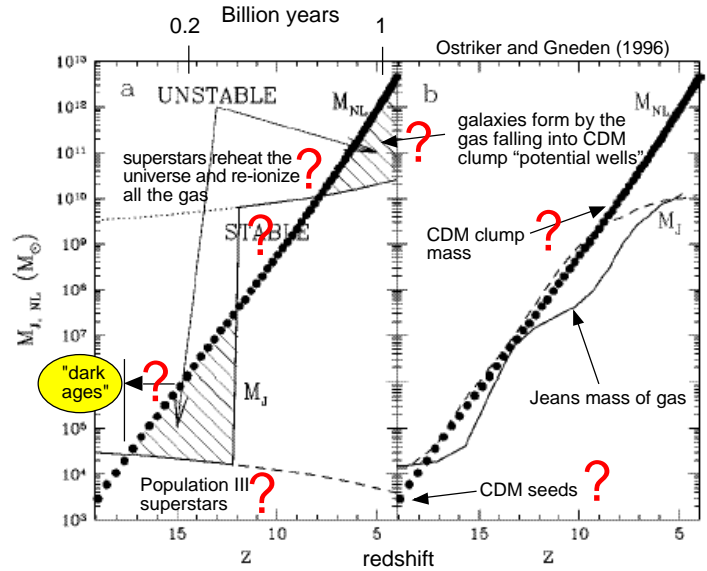

**Figure 2. Reheating of the Universe by Population III superstars in the standard Cold Dark Matter model (Ostriker and Gnedin 1996, fig. 1 L64). In (a) on the left, a few superstars form at redshift z = 17 (175 million years after the big bang) and re-ionize all remaining gas. A stable interval occurs as the CDM seeds clump to dwarf galaxy and galaxy masses between z = 7 and 4 at a billion years when galaxies begin forming. The CDM clump mass ($M_{NL}$) is not much different than the Jeans mass $M_J$ in (b).**

The Ostriker and Gnedin 1996 CDM model shown in Fig. 2 relies on the Press and Schechter 1974 formalism for the self-similar gravitational condensation of self-gravitating masses in an expanding universe. Press-Schechter found by N-body simulations that seed particles become nonlinear and clump to form larger structures in a bottom-up cascade from small to larger masses. They found 6 $10^{37}$ kg seeds at $10^{13}$ s produce a match to superclusters presently observed, larger than the CDM seeds used by Ostriker and Gnedin 1996. All such CDM studies fail to explain how CDM particles can form these sticky seeds, or how "nonlinear" clumps of the CDM seeds can prevent CDM particles from diffusing away. Use of Jeans-CDM makes almost every part of Fig. 2 questionable.

## 3. FLUID MECHANICAL MODEL FOR COSMOLOGICAL STRUCTURES
The Gibson 1996 fluid mechanical model for gravitational structure formation is virtually opposite to the Jeans-CDM model. The largest gravitational structures form first rather than last. The first star appears before the gas is dark, so the "dark ages" of Fig. 2 never happened. The nonbaryonic dark matter is driven to form structures by the baryonic structures, reversing their roles in CDM. Turbulence begins immediately and drives the big bang rather than being suppressed or irrelevant, Gibson 2002. Turbulence in the expanding plasma and gas is not inhibited by the Hubble flow but may be caused by Hubble relative velocities $\gamma L$, where $\gamma$ is the "Hubble constant" $1/t$ and L is a separation distance. Viscosity or buoyancy forces prevent turbulence depending on the Reynolds number $\gamma^2 L/\nu$ and Froude number $\gamma/(G\rho)^{1/2}$ criteria in the usual way. Observations that $\Delta T/T$ in the Cosmic Microwave Background (CMB) is only $10^{-5}$ prove that the primordial plasma flow was non-turbulent because fully developed turbulence driven by the Hubble flow would produce $\Delta T/T$ values orders of magnitude larger; so that either viscosity or buoyancy or both must be damping turbulence in the primordial plasma. Because the viscosity of the primordial plasma near its transition to gas is inadequate to prevent turbulence at $10^{13}$ s, buoyancy forces must be responsible, Gibson 2000. The only way buoyancy forces can arise is if gravitational structure already exists in the primordial plasma, contrary to Jeans' theory. Figure 3 summarizes the major stages of cosmological structure formation in the Gibson 1996, 1997, 2000, 2002 model. Note that Fig. 2 begins at $4 \times 10^{15}$ s, well after Fig. 3. ends.

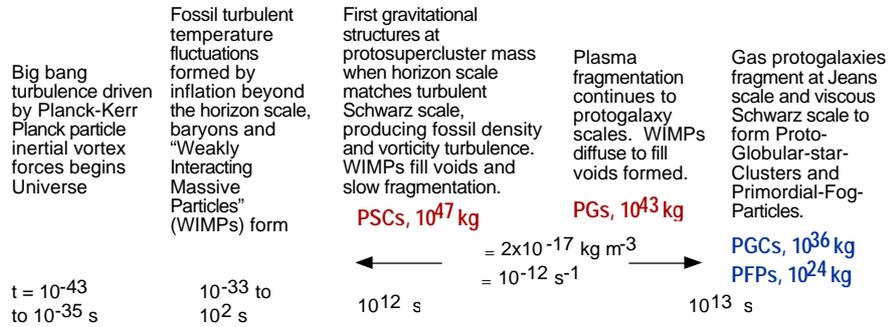

**Figure 3. Cosmological structure formation, Gibson 1996-2002.**

As shown in Fig. 3, no gravitational structure forms for 30,000 years after the big bang as the universe expands and cools with $L_{SV} > L_{ST} > L_H$. Big bang turbulent temperature fluctuations are fossilized by inflation to scales larger than the horizon at $10^{-33}$ s, Gibson 2002. The inflated strong-force freeze-out scale $10^{-2}$ m enters the horizon at t = $10^{-10}$ s, well before nucleosynthesis at $10^2$ s (the first three minutes), Weinberg 1977. Fossil turbulent temperature fluctuations are stretched by the Hubble (uniform straining) flow until the time of nucleosynthesis. Radiation super-viscosities reduce the Reynolds number to subcritical values $< 10^2$, preventing any turbulence until late in the plasma epoch after matter dominates radiation at $10^{10}$ s. Nucleosynthesis rate constants are extremely sensitive to temperature (~ $T^5$, Peacock 2000) so fossil temperature turbulence fluctuations persisting at nucleosynthesis can produce hydrogen-helium plasma density fluctuations (i.e., fossil density turbulence) to trigger the first gravitational formation of structure at about $10^{12}$ s.

Gibson 1997 shows the horizon mass $\rho(t)[ct]^3$ matches the supercluster mass at time $10^{12}$ s, Figure 4, where $\rho(t)$ is the critical density for a flat universe and t is the time since the big bang, calculated using Einstein's equations in Weinberg (1972, p540). The supercluster mass $10^{47}$ kg is $10^{-6}$ times the present horizon mass $10^{53}$ kg because the supercluster scale is observed to be $10^{-2}$ times the horizon scale. Fig. 4 is redrawn from Fig. 3 in Gibson 1997.

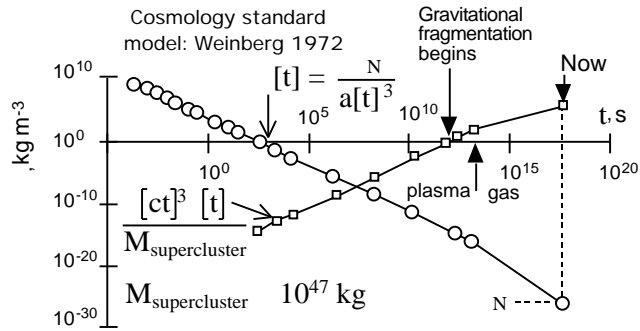

**Figure 4. Hydropaleontology using the observed supercluster mass: the horizon-mass/supercluster-mass ratio increases with time t and the critical density decreases (from Einstein's equations), Weinberg (1972, p540), suggesting the time of the first gravitational structure formation was $10^{12}$ seconds (30,000 years), Gibson 1997.**

The cosmological scale factor a(t) in Fig. 4 is the ratio $R(t)/R_o$, where R(t) is the size a length $R_o$ at present was at time t before space was stretched according to the general relativity equations of Einstein. For example, at time $2 \times 10^{12}$ s, a(t) was $2.7 \times 10^{-4}$, Weinberg 1972, p540, so the critical density then was $10^{-26}/a^3 = 5 \times 10^{-16}$ kg m$^{-3}$. Most of the matter was nonbaryonic, by a factor of about 30 to 1, so the baryonic matter density at the time of first fragmentation was about $\rho_o = 2 \times 10^{-17}$ kg m$^{-3}$, as shown in Fig. 3. This initial density of fragmentation should be preserved, along with the rate-of-strain $\gamma_o = 10^{-12}$ s$^{-1}$, as a fossil of this event, as shown by arrows in Fig. 3. This fossil value $\rho_o$ is the density of all PSCs, PGs, PGSs, and PFPs formed in Fig. 3. $\rho_o$ is preserved in many fossils of this first fragmentation such as the size of galaxy cores, about $10^{20}$ m, and the core densities of globular star clusters. Figure 5 shows a recent Hubble Space Telescope image of the core of the Milky Way globular star cluster Omega Centauri (NGC 5139), for comparison.

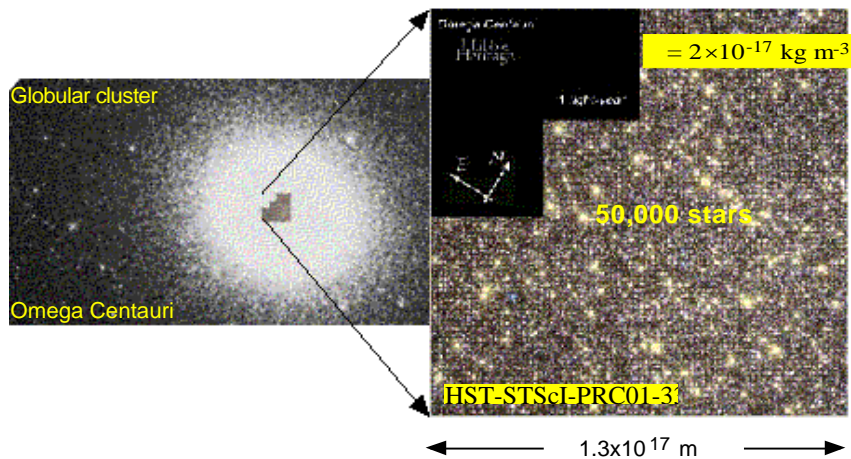

**Figure 5. Globular star cluster Omega Centauri core imaged by the HST.**

The ω Cen core density is $2 \times 10^{-17}$ kg m$^{-3}$ from the observed 50,000 stars in the $1.3 \times 10^{17}$ m HST core image, matching the fragmentation baryon density of the universe at $10^{12}$ s according to the scenario of Fig. 3. The total number of stars in the globular cluster is over $2 \times 10^6$ making ω Cen the largest globular cluster in the Galaxy in terms of numbers of stars. Most of the other 200 Milky Way globular clusters have similar core densities and star numbers with ages of 13.5±0.5 Gy, Ashman and Zepf 1998. Such large globular cluster ages coincide with recent best estimates for the age of the universe, and seem to rule out the billion years (1 Gy) of CDM clustering and galaxy formation shown in Fig. 2.

Bright globular star clusters in all galaxies all have nearly the same mass, size, age and density, which is not easily (or possibly) explained by the CDM scenario of Fig. 2. The initial violent formation of Population III superstars in Fig. 2 not only reionizes the gas but makes the Jeans mass too large to permit the uniform mass near $10^6$ M$_{SUN}$ of bright globular star clusters observed. In the Gibson 1996 scenario of Fig. 3, formation of primordial fog particles prevents the formation of Population III superstars (or any stars) in the initial stages of PGC fragmentation (except at the cores of proto-galaxies). Most stars are formed by an accretional cascade of PFPs, and most PFPs have never formed stars but persist as the baryonic dark matter. [The Pop. III exception is that the increased probability of collisions at proto-galaxy centers of mass might cause a maelstrom of turbulent gas with large enough turbulent Schwarz scales L$_{ST}$ and sufficient cooling to prevent star formation until the rising density permits collapse to a superstar, a supernova, a black hole and a γ-ray burst]. Many central black holes have been identified in galaxies, but none (so far) at the center of globular star clusters. Several dim globular clusters such as Polomar 13 have been detected in the Milky Way far out in the halo, at $10^{21}$ m or more, with only $10^3$ to $10^4$ stars concentrated in small cores. Most of the mass of such globulars may be dark and frozen as PFPs that are bound by gravity and friction. The high resolution of the HST has revealed numerous young globular star clusters in starburst or merging galaxies such as NGC 6946, 1705, NGC 1569, the Antennae, NGC 7252 and NGC 3256, with ages of only $10^6$ years or less, Larsen et al. 2001. It seems likely that these young globulars were formed from dark PGCs and their unaccreted PFPs. Population III stars rarely form in the scenario of Fig. 3. No re-ionization of the universe is needed to explain the Gunn-Peterson missing neutral gas because the hydrogen-helium is not uniformly distributed as in Fig. 2, but is condensed out of the line of sight of quasars in the form of frozen planets starting about redshift 30 when the temperature of the background radiation falls below the freezing point of hydrogen.

The mass of primordial-fog-particles is determined by the viscous Schwarz scale $L_{SV} = (νγ/G\rho)^{1/2}$ existing in the proto-galaxy gas blobs emerging from the plasma epoch. The density should be the fossil density turbulence value $\rho_o = 2 \times 10^{-17}$ kg m$^{-3}$ from the first fragmentation time $10^{12}$ s, as shown in Fig. 3, and the rate of strain γ should be the fossil vorticity turbulence value $\gamma_o = 10^{-12}$ s$^{-1}$ (ε ≈ 200 m$^2$ s$^{-3}$). Fossil turbulence formation and properties are reviewed by Gibson 1999. The weighted average hydrogen-helium gas viscosity at 3000 K is $2.4 \times 10^{-4}$ kg m$^{-1}$ s$^{-1}$, so ν is $1.3 \times 10^{13}$ m$^2$ s$^{-1}$, giving $L_{SV} = 9.8 \times 10^{13}$ m and $M_{PFP} = \rho L_{SV}^3 = 1.9 \times 10^{25}$ kg, three times the mass of the Earth, which is an upper bound since both the density and rate-of-strain may have been reduced between $10^{12}$ and $10^{13}$ s by expansion or viscous friction in the proto-galaxies. If the assumed density is $10^{-17}$ and the assumed γ is $10^{-13}$, then M$_{PFP}$ is reduced to $2.1 \times 10^{24}$ kg, a third of the mass of the Earth.

10/12/01 8:21 AM 9

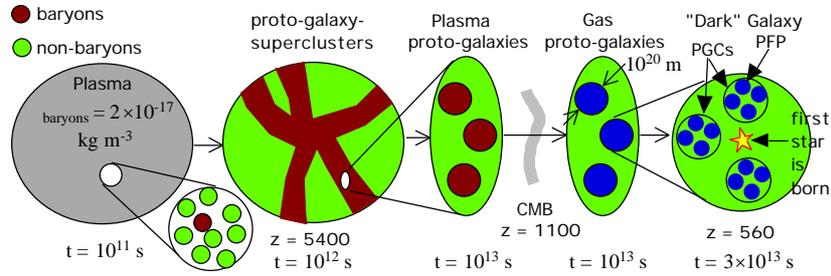

**Figure 6. The first structure formation from PSCs to "Dark" Galaxies**

Figure 6 shows the first gravitational structure formation of Fig. 3. The progression of structure formation is from left to right in time starting with a mix of baryonic and non-baryonic particles of the hot plasma before formation of PSC foam with non-baryonic particles diffusing to fill the voids. "Dark" galaxies result with 100% dark (i. e., invisible) matter, with the primordial hydrogen-helium gas fragmented at the Jeans scale to form $10^{36}$ kg PGC clumps of $10^{25}$ kg PFPs and the first stars in the free fall time $_{FF} = 3 \; 10^{13}$ s for  = $2 \; 10^{-17}$ kg m$^{-3}$. At 1000 K when the first star was born, the Universe glowed like molten aluminum. The "dark ages" of Fig. 2 never happened in Fig. 6.

**4. OBSERVATIONS**
Evidence for the existence of primordial fog particles is provided by photographs of the Helix Planetary Nebula by O'Dell. using the Hubble Space Telescope as shown in Figure 7.

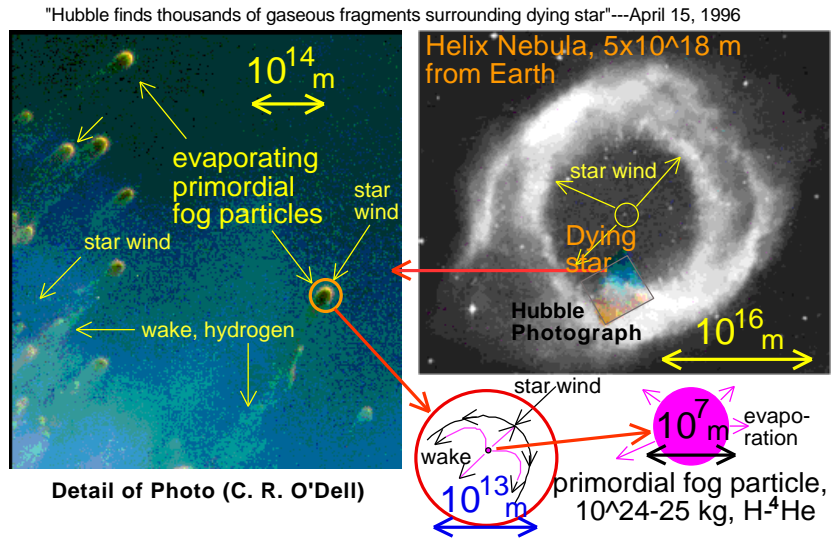

**Figure 7. HST photographs of PFP candidates in the Helix planetary nebula.**

Planetary nebulae are formed when a solar mass star exhausts its hydrogen and helium fuel and is left with nothing but its extremely hot carbon core, which will eventually turn into a white dwarf. Ultraviolet radiation and plasma jets emitted by the dying star illuminate the surrounding interstellar medium at distances about $10^{16}$ m from the star, creating a glowing region that looked like a planet through small telescopes. Helix is the closest planetary nebula to the earth at $5\ 10^{18}$ m, so cometary globules only dimly visible from surface telescopes gave spectacular images in the HST photographs of C. R. O'Dell shown in Fig. 7. The mass of some of the larger objects is reported to be about $10^{25}$ kg, supporting the Fig. 7 interpretation that these are $10^{13}$ m photo-ionized cocoons with PFPs inside that have been partially evaporated by radiation and wind from the central dying star. Little evidence supports the claim by some authors that the gas and cometary knots of planetary nebulae are ejected from the central star, rather than emerging out of the dark by evaporation of PFPs that dominate the mass of the interstellar medium as the baryonic dark matter.

Further evidence that planetary mass objects dominate the mass of galaxies is provided by the dominant twinkling frequency of background quasar images lensed by a foreground galaxy, such as the twinkle in Figure 8 reproduced from Schild 1996.

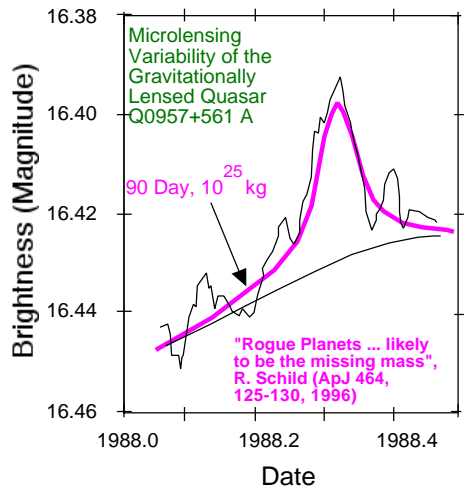

**Figure 8. Light curve of a lensed quasar image, Schild 1996.**

As shown in Fig. 8, Schild 1996 independently reaches the same conclusion as Gibson 1996; that is, that the missing mass of galaxies is in the form of small-planet-mass objects.

## 5. CONCLUSIONS
Jeans-Cold-Dark-Matter scenarios for gravitational structure formation in cosmology and astrophysics are highly questionable. The Jeans 1902 theory and length scale criterion for gravitational instability neglects viscous, turbulence and diffusional effects that are all important to the formation of the first gravitational structures. A variety of fluid mechanical misconceptions associated with the CDM concept are identified. Because CDM particles are collisionless they cannot cool, condense, or clump as assumed, and cannot guide gravitational structure formation by providing the gravitational potential wells needed to salvage the Jeans 1902 criterion by CDM. Schwarz-length-scale criteria predict gravitational structure formation in better agreement with observations of early galaxy existence and clustering and resolve both the dark matter paradox and the Gunn-Peterson missing-gas paradox. Population III superstars never form in great numbers and the universe never re-ionizes in the Gibson 1996-2002 scenario. Evidence that the baryonic dark matter consists of proto-globular-cluster PGC clumps of frozen hydrogen-helium primordial-fog-particle PFP objects is provided by observations of young-globular-clusters (YGCs) formed in starburst and merger galaxies. YGC densities and masses match the old-globular values, suggesting that YGCs are formed from ancient dark PGCs. A large number of PFPs are expected in the interstellar medium of the Galaxy disk from PGCs disrupted by tidal forces as they pass through. Such unattached PFPs may be detected near hot objects by their large luminous photo-ionized evaporation shells, as shown in the Helix planetary nebula of Fig. 7. Globular star clusters are observed in galaxies at radial distances 3 $10^{21}$ m, suggesting the Dark-PGC-PFP (DPP) baryonic dark matter halo probably extends to greater distances. This may explain why the MACHO and EROS collaborations failed to detect any earth-mass objects microlensing stars of the Large Magellanic Cloud 1.6 $10^{21}$ m from the earth. Most of the Milky Way DPP dark matter may be far beyond the LMC stars, along with the nonbaryonic dark matter halo at $10^{22}$ m.

**ACKNOWLEDGEMENTS**
The author would like to thank Professor Fazle Hussain for discussions and comments leading to this paper.

**REFERENCES**
Ashman, K. M. and S. E. Zepf (1998), *Globular Cluster Systems*, Cambridge University Press, UK.

Binney, J. and S. Tremaine (1987), *Galactic Dynamics*, Princeton University Press, New Jersey.

Ciardi, B.; Ferrara, A.; Governato, F.; Jenkins, A. (2000), "Inhomogeneous reionization of the intergalactic medium regulated by radiative and stellar feedbacks", *Monthly Notices of the Royal Astronomical Society*, vol.314, (no.3):611-29.

Gibson, C. H. (1996), "Turbulence in the ocean, atmosphere, galaxy, and universe", *Applied Mechanics Reviews*, 49:5, 299-315.

Gibson, C. H. (1997), "Dark matter at viscous-gravitational Schwarz scales: theory and observations", *Dark Matter in Astro- and Particle Physics*, Eds. H. V. Klapdor-Kleingrothaus and Y. Ramachers, World Scientific, Singapore, 409-416.

Gibson, C. H. (1999), "Fossil turbulence revisited", *Journal of Marine Systems*, vol. 21, nos. 1-4, 147-167.

10/12/01 8:21 AM  12Gibson, C. H. (2000), "Turbulent mixing, diffusion and gravity in the formation of cosmological structures: the fluid mechanics of dark matter", *Journal of Fluids Engineering*, 122, 830-835.

Gibson, C. H. (2002), "Turbulence and Mixing in the Early Universe", *Conference Proceedings*, 4th International Conference on Mechanical Engineering, Dhaka, 2001.

Jeans, J. H. (1902), "The stability of a spherical nebula", *Phil. Trans. R.. Soc. Lond. A*, 199, 1.

Kolb, E. W. and M. S. Turner (1990), *The Early Universe*, Addison Wesley, NY.

Larsen, S. S., J. P. Brodie, B. G. Elmegreen, Y. N. Efremov, P. W. Hodge, and T. Richtler (2001), "Structure and mass of a young globular cluster in NGC 6946", *The Astrophysical Journal*, 556, 801-812.

Ostriker, J. P. and N. Y. Gnedin (1996), "Reheating of the universe and population III", *Astrophysical Journal*, vol. 472, L63-L67.

Peacock, J. A. (2000), *Cosmological Physics*, Cambridge University Press.

Padmanabhan, T. (1993), *Structure Formation in the Universe*, Cambridge University Press, Cambridge UK.

Peebles, P. J. E. (1993), *Principles of Physical Cosmology*, Princeton University Press, Princeton, NJ.

Peebles, P. J. E. (2001), "The void phenomenon", *The Astrophysical Journal*, 557:495-504.

Press, W. H. and P. Schechter (1974), "Formation of galaxies and clusters of galaxies by self-similar gravitational condensation", *Astrophysical Journal*, vol. 187, 425-438.

Rees, Martin (2000), *New Perspectives in Astrophysical Cosmology*, Cambridge University Press, UK.

Schild, R. E. (1996), "Microlensing variability of the gravitationally lensed quasar Q0957+561 A,B", *Astrophys. J.*, 464, 125-130.

Silk, Joseph (1989), *The Big Bang*, W. H. Freeman and Company, NY.

Steidel, C.C.; Adelberger, K.L.; Shapley, A.E.; Pettini, M.; and others (2000), "Ly alpha imaging of a proto-cluster region at <z>=3.09", *Astrophysical Journal*, vol.532, (no.1, pt.1):170-82.

Weinberg, S. (1972), *Gravitation and Cosmology: Principles and Applications of the General Theory of Relativity*, John Wiley & Sons, New York.

Weinberg, S. (1977), *The First Three Minutes*, Basic Books, Inc., Publishers, New York.
12